\documentclass[aps,prd,twocolumn,showpacs,floatfix,nofootinbib,superscriptaddress]{revtex4-1}
\usepackage{amsmath}
\usepackage{amssymb}
\usepackage{graphicx}
\usepackage{xurl}
\usepackage{hyperref}
\usepackage{color}
\usepackage{mathtools}
\usepackage{times}
\usepackage{bm}
\usepackage{xcolor}
\usepackage{comment}

\usepackage[normalem]{ulem}

\newcommand{\be}{\begin{equation}}
\newcommand{\ee}{\end{equation}}
\newcommand{\beq}{\begin{equation}}
\newcommand{\eeq}{\end{equation}}
\newcommand{\bea}{\begin{eqnarray}}
\newcommand{\eea}{\end{eqnarray}}

\newcommand{\de}{\delta}
\newcommand{\ep}{\epsilon}

\newcommand{\lam}{\lambda}
\newcommand{\Lam}{\Lambda}

\newcommand{\obs}{{\rm o}}
\newcommand{\emi}{{\rm e}}

\renewcommand{\be}{\beta}

\newcommand{\bean}{\begin{eqnarray*}}
\newcommand{\eean}{\end{eqnarray*}}

\newcommand{\id}{{\rm 1\kern -2.5pt I}}

\newcommand{\znu}{z_\nu}
\newcommand{\ztau}{z_\tau}

\begin{document}

\title{Exact Equivalence of the Observed Redshift and the Pulsar Timing Modulation in the Infinitesimal-Pulse Limit}

\author{Matteo Magi}
\email{mmagi@ibs.re.kr}
\affiliation{Cosmology, Gravity and Astroparticle Physics Group,
 Center for Theoretical Physics of the Universe, Institute for 
Basic Science (IBS), Daejeon, 34126, Korea}
\author{Jaiyul Yoo}
\email{jaiyul.yoo@uzh.ch}
\affiliation{Center for Theoretical Astrophysics and Cosmology,
Department of Astrophysics,
University of Zurich, Winterthurerstrasse 190,
CH-8057, Zurich, Switzerland}
\affiliation{Department of Physics, University of Zurich,
Winterthurerstrasse 190, CH-8057, Zurich, Switzerland}

\date{\today}
\begin{abstract}
The pulsar timing arrays (PTA) collect the times of arrival of radio signals from the millisecond pulsars, and gravitational waves can modulate their arrival times. Although the PTA observable involves
successive radio pulses propagating along different light paths, its standard theoretical description at linear order in perturbations is equivalent to the Sachs--Wolfe formula for the observed redshift, which describes the fractional change in photon frequency along a single geodesic. While this equivalence is well known at linear order, its validity beyond first order has not been systematically addressed in the PTA literature. Here we show that, in the limit of vanishing proper-time separation between successive emission events, the timing modulation is
exactly equal to the observed redshift, without expanding the spacetime geometry. For a finite emission interval, we derive the exact relation between the timing modulation and the observed redshift, and show that their
difference is controlled by the ratio between the emission interval and the characteristic timescale over which the observed redshift varies.
\end{abstract}

\maketitle

\section{Introduction}

With the discovery of stable millisecond pulsars, pulsar timing arrays (PTA) can precisely measure the times of arrival of radio pulses from individual pulsars over a long period of time~\cite{BAKUET82,FOBA90}, and these PTA data can be used to look for possible sources such as gravitational waves that can further modulate the times of arrival of radio pulses during their propagation from the pulsars~\cite{SAZHI78,DETWE79,HEDO83,WITTE84,NANO23new,NANO23gw,EPTA23C,EPTA23d} (see, e.g.,~\cite{BAHE86,CAFI18,MAGGI18} for reviews). Many PTA collaborations~\cite{IPTA10,NANO13,KRCH13,PPTA13} around the globe have accumulated precise timing data over decades, and tantalizing evidence for gravitational waves in the angular cross-correlation of the pulsar timing residuals, known as the Hellings--Downs curve~\cite{HEDO83}, was also found~\cite{NANO23gw,EPTA23C,PPTA23gw,CPTA23gw}.

The standard theoretical description developed in the pioneering works of Refs.~\cite{ESWA75,SAZHI78,DETWE79} shows that, at linear order in perturbations, the modulation of pulsar arrival times by external sources, such as gravitational waves or scalar fields, is described by the Sachs--Wolfe formula~\cite{SAWO67} for the observed redshift specialized to a Minkowski background. A light source emits electromagnetic waves in its rest frame with frequency~$\nu_\emi$, and an observer measures the signals with frequency~$\nu_\obs$ in their rest frame. The observed redshift~$z_\nu$ is then defined as
\begin{equation}
  1+z_\nu \equiv \frac{\nu_\emi}{\nu_\obs}
  =\frac{(-u^\mu k_\mu)_\emi}{(-u^\mu k_\mu)_\obs},
  \label{z1}
\end{equation}
where~$k^\mu$ is the physical photon wave-vector and~$u^\mu$ is the four-velocity of the emitter~($\emi$) and the observer~($\obs)$, defining their rest-frames. The redshift~$z_\nu$ describes how the photon frequency (or wavelength) emitted in the rest frame of the source is modulated along the light propagation until it is measured in the rest frame of the observer. The linear-order description of the observed redshift~$z_\nu$ in Eq.~\eqref{z1} is the Sachs--Wolfe formula~\cite{SAWO67}, and it is widely used in cosmology (see, e.g.,~\cite{WEINB72,WEINB08B,PEACO99,DODEL03}).

In contrast, the PTA measures the times of arrival of radio signals from millisecond pulsars. These pulsars emit a sequence of radio signals at millisecond intervals in their rest frame. In the idealized description adopted here, we denote by~$\Delta\tau_\emi$ the proper-time interval between two consecutive emission events. An observer at the PTA records the arrival times of these signals to obtain the observed proper-time interval~$\Delta\tau_\obs$ between two consecutive arrival times, which can be used to define the observed timing modulation~$z_\tau$ as
\begin{equation}
  1+z_\tau \equiv \frac{\Delta\tau_\obs}{\Delta\tau_\emi},
  \label{z2}
\end{equation}
which describes how the proper-time interval~$\Delta\tau_\emi$ in the pulsar rest frame is modulated along the light propagation of two consecutive signals until they are measured in the rest frame of the observer. In the literature~\cite{ANBAET09,QIBOET19,NANOnonGR21,DOPIWA25},~$z_\tau$ is often referred to as the observed redshift, albeit not the main convention.

It is clear that these two redshifts~$z_\nu$ and~$z_\tau$ are physically different. The former~$z_\nu$ describes the physical change of the photon frequency (or wavelength) along a single light path, and it requires a spectroscopic instrument to measure the redshift~$z_\nu$. The latter~$z_\tau$ describes the arrival times (or arrival frequency) of two consecutive radio signals along two different light paths, and no spectroscopic instrument is needed for the measurements. Furthermore, the physical frequency of emitted signals from a pulsar (whether the pulsar emits radio signals or X-ray signals) is irrelevant for the measurement of~$z_\tau$.

Hence, while the equivalence of the two redshifts~$z_\nu$ and~$z_\tau$ is well known at linear order in perturbations, its validity beyond
first order has not been systematically addressed in the PTA literature. The natural questions arise: Is this equivalence valid beyond linear order and beyond general relativity, and what is the reason behind it? Although related geometrical results concerning local variations of light propagation have long been known in the literature~\cite{perlick2000ray,Straumann:2013spu}, their relevance to the PTA timing observable beyond linear order has not been made explicit. Here we derive the equivalence directly from the two-pulse timing observable, carefully accounting for the neighboring null geodesics and their boundary conditions. We show the exact nonlinear equivalence of the two redshifts in any metric theory in the limit of vanishing proper-time interval~$\Delta\tau_\emi$, and derive their corresponding relation for a finite emission interval.

\section{Geometrical construction}

We now introduce the geometrical framework needed to relate the two different redshifts~$\znu$ and~$\ztau$ in Eqs.~\eqref{z1} and~\eqref{z2}. Throughout this work, we assume that electromagnetic signals propagate along null geodesics of a given metric, with each emission event connected to a unique observation event by such a geodesic. The source and observer follow arbitrary timelike worldlines parametrized by their proper times, and their motions need not be geodesics. Our analysis here is purely kinematical and applies to any metric theory
of gravity.

We model a millisecond pulsar as a source clock emitting a sequence of radio signals. Let~$x^\mu(\lambda)$ denote the null geodesic of the first radio pulse, parametrized by an affine parameter~$\lambda$, connecting the emission of the first radio pulse at the source emission proper time~$\tau_{\emi}$ to its observation at the observer proper time~$\tau_{\obs}$. Similarly, let~$x^\mu(\lambda')$ denote the null geodesic of the second radio pulse, parametrized by another independent affine parameter~$\lambda'$, connecting the emission of the second pulse at~$\tau'_{\emi}$ to its observation at~$\tau'_{\obs}$. We set $\lam_{\obs}=0=\lam'_{\obs}$ at the corresponding observation events $\tau_\obs$ and $\tau'_\obs$. Fixing the affine origins removes the freedom to translate~$\lambda$ and~$\lambda'$.  The remaining affine freedom is a constant rescaling of each parameter, which may be fixed, for example, by requiring the corresponding tangent vector~$k^\mu=dx^\mu/d\lambda$ to coincide with the physical photon wave vector at the observation event~\cite{YOGRET18}. With this convention, the affine values of the two emission events generally differ, and we write $\lam'_{\emi}\equiv\lam_{\emi}+\delta\lam_{\emi}\,$.

To describe the light propagation of two neighboring radio pulses, we consider the continuous-emission limit, in which $\Delta\tau_{\emi}\to0$. In this limit, the corresponding null geodesics can be represented as a smooth one-parameter family, $x^\mu=x^\mu(\Lambda,\epsilon)$, where the parameter~$\epsilon$ labels different geodesic paths, while~$\Lambda$ is an affine parameter along each null geodesic. We choose~$\epsilon=0$ to identify the reference ray associated with the first radio pulse, and represent the second radio pulse by a neighboring member of the geodesic family separated by an infinitesimal parameter interval~$d\epsilon$. On the reference and neighboring rays,~$\Lambda$ reduces to~$\lambda$ and~$\lambda'$,
respectively. The affine parametrization is chosen such that~$\Lam=0$ at the corresponding observation event for every member of the family, while the affine parameter at emission is allowed to vary.

Along the reference ray, we introduce the null tangent vector~$k^\mu$ and the infinitesimal displacement field~$X^\mu$
\begin{equation}
    k^\mu \equiv \left. \frac{\partial x^\mu}{\partial\Lam} \right|_{\ep=0}\,,
    \qquad\qquad
    X^\mu \equiv d\ep \left. \frac{\partial x^\mu}{\partial\ep} \right|_{\ep=0}\,.
\end{equation}
The vector~$k^\mu(\Lambda)$ follows the reference geodesic path from emission to observation, while~$X^\mu(\Lambda)$ connects each point of the reference ray to the corresponding point of the neighboring ray at the same affine parameter~$\Lambda$, representing the infinitesimal separation between two neighboring rays associated with successive radio pulses.

Since~$X^\mu$ is the separation field associated with a smooth family of null geodesics, it is a Jacobi field and satisfies the geodesic-deviation equation (see, e.g.,~\cite{WEINB72}),
\begin{equation}
    \frac{D^2 X^\mu}{d\Lam^2} = R^\mu{}_{\nu\rho\sigma} k^\nu k^\rho X^\sigma\,,
    \qquad\qquad
    \frac{D}{d\Lam}\equiv k^\mu\nabla_\mu\,.
\end{equation}
The geodesic-deviation equation describes how the separation between neighboring null rays evolves along their propagation. The transverse components of~$X^\mu$ govern the relative position and direction of neighboring rays, and therefore underlie astrometric and lensing observables~\cite{YOGRET18,GRYO18,Magi:2026xcy}. Here, however, we only need
the scalar contraction~$k_\mu X^\mu$.

\section{Exact nonlinear equivalence of two redshifts}
\label{sec:proof}

The key ingredient in comparing two redshifts~$\znu$ and~$\ztau$ in Eqs.~\eqref{z1} and~\eqref{z2} is the limit of vanishing proper-time interval for consecutive radio pulses. In this limit, the neighboring null geodesic differs from the reference ray by the displacement field~$X^\mu$, up to corrections of order~$\mathcal{O}(d\epsilon^2)$. Since~$\Lambda$ and~$\epsilon$ are coordinates on the two-dimensional surface swept out by the geodesic congruence, the corresponding vector fields commute on this surface, and the Lie derivative~${\cal L}_kX=0$:
\begin{equation}
    k^\nu\nabla_\nu X^\mu = X^\nu\nabla_\nu k^\mu\,.
\end{equation}
This identity implies that the scalar~$k_\mu X^\mu$ is conserved along the reference geodesic. Indeed,
\begin{align}
\label{main}
    k^\nu\nabla_\nu\left(k_\mu X^\mu\right)
    &=
    \left(k^\nu\nabla_\nu k_\mu\right)X^\mu + k_\mu k^\nu\nabla_\nu X^\mu
    \nonumber\\
    &=
    k_\mu X^\nu\nabla_\nu k^\mu = \frac{1}{2}X^\nu\nabla_\nu
    \left(k^\mu k_\mu\right)
    =
    0\,,
\end{align}
where we used the affinely parametrized geodesic equation ($0=k^\nu\nabla_\nu k^\mu$) and the null condition~($k^\mu k_\mu=0$). Hence, we obtain
\begin{equation}\label{kXcons}
    \left(k_\mu X^\mu\right)_{\obs} = \left(k_\mu X^\mu\right)_{\emi}\,.
\end{equation}

When the family of null rays is generated by a common smooth phase field~$\varphi(x^\mu)$, with
$k^\mu=\nabla^\mu\varphi$, the conserved scalar~$k_\mu X^\mu$ also admits a phase interpretation. The phase is constant along the reference geodesic as a consequence of the null condition:
\begin{equation}\label{phiconst}
    \frac{d\varphi}{d\Lam} = k^\mu\nabla_\mu\varphi = k^\mu k_\mu = 0\,.
\end{equation}
Since the phase difference between two neighboring rays at fixed common affine parameter~$\Lam$ is
\begin{equation}
    \de\varphi = X^\mu\nabla_\mu\varphi = k_\mu X^\mu\,,
\end{equation}
Eq.~\eqref{kXcons} is therefore equivalent to stating that the phase difference between neighboring signals is independent of~$\Lam$, and hence has the same value at emission and observation. This provides a useful interpretation of the conservation law, although the derivation below requires only the conservation of~$k_\mu X^\mu$.

We now evaluate the conserved scalar at the emission and observation events by relating the displacement field to the source and observer four-velocities~$u^\mu$. The second radio pulse is emitted at $\tau'_{\emi} = \tau_{\emi}+d\tau_{\emi}\,$ and observed at $\tau'_{\obs} = \tau_{\obs}+d\tau_{\obs}\,$. At observation, where the affine origins of the two geodesics coincide ($\lam_\obs=\lam'_\obs=0$), the displacement~$X^\mu_\obs$ at fixed affine parameter coincides with the physical separation between the two observation events, \begin{equation}\label{Xobs}
    u^\mu_{\obs}\,d\tau_{\obs}=X^\mu_\obs\,,
\end{equation}
where we have used that~$d\tau_\obs$ is infinitesimally small. At emission, the two rays do not intersect the source worldline at the same value of their affine parameters. The physical separation between the emission events therefore contains both the displacement at fixed affine parameter and the shift of the affine endpoint,
\begin{equation}\label{Xemi}
    u^\mu_{\emi}\,d\tau_{\emi} = X^\mu_{\emi} + k^\mu_{\emi}\,\delta\lambda_{\emi}\,.
\end{equation}
Contracting Eqs.~\eqref{Xobs} and~\eqref{Xemi} with~$k_\mu$ and using $k^\mu k_\mu=0$ yields
\begin{equation}
    \left(k_\mu X^\mu\right)_{\obs} = \left(k_\mu u^\mu\right)_{\obs}d\tau_{\obs}\,,
    \qquad\quad
    \left(k_\mu X^\mu\right)_{\emi} = \left(k_\mu u^\mu\right)_{\emi}d\tau_{\emi}\,.
\end{equation}
The conservation law in Eq.~\eqref{kXcons} then implies
\begin{equation}\label{mainidentity}
    \lim_{\Delta\tau_{\emi}\to0}\left(1+\ztau\right)
    =
    \frac{d\tau_{\obs}}{d\tau_{\emi}}
    =
    \frac{\left(k_\mu u^\mu\right)_{\emi}}
         {\left(k_\mu u^\mu\right)_{\obs}}
    =
    \frac{\nu_{\emi}}{\nu_{\obs}}
    =
    1+\znu\,.
\end{equation}
This proves that, in the limit of vanishing proper-time interval $\Delta\tau_\emi=d\tau_\emi\rightarrow0$ between two consecutive radio pulses, the observed pulsar timing modulation~$\ztau$ approaches the observed redshift~$\znu$.

The geometrical origin of this exact equivalence is the conservation of~$k_\mu X^\mu$ along the reference ray. When the family of rays is generated by a common phase field, this conservation law can be interpreted as conservation of the phase difference~$\delta\varphi$ between neighboring rays, which yields the relation $-\delta\varphi=\nu_{\emi}d\tau_{\emi}=\nu_{\obs}d\tau_{\obs}$, i.e., the same infinitesimal phase difference between two consecutive signals
is therefore present at emission and observation.
At the emission, the frequency~$\nu_\emi$ of the radio pulses and the proper-time interval~$d\tau_\emi$ (pulsar rotational period) determine the phase 
difference~$\delta\varphi$. Once $\delta\varphi$ is determined, 
subsequent changes
of frequency~$\nu$ by the spacetime fluctuations during the light propagation
induce the change in the proper-time interval~$d\tau$  to keep the same
phase difference~$\delta\varphi$. At the observer, the fractional change
in the observed frequency~$\nu_\obs$ is, therefore,
 inversely proportional to the
fractional change in the proper-time interval~$d\tau_\obs$.
This exact equivalence between~$z_\tau$ and~$z_\nu$ holds in the infinitesimal
limit. However, note 
 that the emission frequency~$\nu_\emi$ (which can be radio or X-ray)
and the emission interval $d\tau_\emi$ (pulsar rotation period)
are independent, and
the argument that the emission frequency~$\nu_\emi$ works as an inverse of the proper-time interval~$d\tau_\emi\sim1/\nu_\emi$ is {\it not} the correct
reason for the equivalence between~$z_\tau$ and~$z_\nu$.

\section{Finite-period corrections}

For a finite proper-time interval~$\Delta\tau_\emi$, the differential relation derived in the previous section can be integrated along the source worldline. To this end, we introduce a smooth function~$\mathcal{T}$, which associates each emission event~$\tau_\emi$ on the source worldline with the unique corresponding observation event $\tau_{\obs}=\mathcal{T}(\tau_{\emi})$. The second radio pulse emitted after a proper-time interval~$\Delta\tau_{\emi}$ is therefore received after the proper-time interval
\begin{equation}
    \Delta\tau_{\obs} = \mathcal{T}(\tau_{\emi}+\Delta\tau_{\emi}) - \mathcal{T}(\tau_{\emi})\,.
\end{equation}
The observed timing modulation~$\ztau$ of the PTA signals in Eq.~\eqref{z2} can thus be written as
\begin{equation}\label{finite-map}
    1+\ztau = \frac{ \mathcal{T}(\tau_{\emi}+\Delta\tau_{\emi}) - \mathcal{T}(\tau_{\emi})}{\Delta\tau_{\emi}}\,.
\end{equation}
Taking the limit~$\Delta \tau_\emi\to0$ and using Eq.~\eqref{mainidentity}, we obtain
$d\mathcal{T}/d\tau_\emi=1+\znu\,$,
so that the second pulse in this limit serves only to define the local variation of the emission--observation map.
At finite proper-time interval~$\Delta\tau_\emi$, the single-ray redshift~$\znu$ generally varies between the two emission events, and the observed timing modulation~$\ztau$ is then
\begin{equation}\label{finite-average}
    1+\ztau = \frac{1}{\Delta\tau_{\emi}}\int_{\tau_{\emi}}^{\tau_{\emi}+\Delta\tau_{\emi}} d\tilde\tau_{\emi}\,
    \Big[1+\znu(\tilde\tau_{\emi})\Big]\,,
\end{equation}
where $\znu(\tilde\tau_\emi)$ is evaluated along the null geodesic emitted at~$\tilde\tau_\emi$.

For a sufficiently smooth~$\znu(\tau_\emi)$, the exact expression may be expanded in powers of the proper-time
interval~$\Delta\tau_{\emi}$ between successive radio pulses,
\begin{equation}\label{finite-expansion}
    \ztau = \znu + \frac{\Delta\tau_{\emi}}{2} \frac{d \znu}{d\tau_{\emi}} +\mathcal{O}\!\left({\Delta\tau_{\emi}^2\over T_z^2}\right)\,,
\end{equation}
where all quantities on the right-hand side are evaluated at the first emission event. The accuracy of the truncated expansion in Eq.~\eqref{finite-expansion} is controlled by the timescale~$T_z$, on which~$\znu$ varies, and the leading finite-period correction to the observed redshift~$\znu$ is suppressed by the parameter $\Delta\tau_\emi/T_z$. Here, $\Delta\tau_\emi$ denotes the actual proper-time separation between the two emission events in the pulsar rest frame, which need not coincide with the pulsar rotation period defined before accounting for the perturbation of the emission events. For a monochromatic gravitational wave with angular frequency~$\omega$, $\znu$ varies on a timescale of order~$1/\omega$, so the expansion is controlled by~$\omega\Delta\tau_\emi$. Finite-period effects have also been considered in the literature, for example in Ref.~\cite{BOTIET20}.

In PTA observations, the pulsar rotation period is typically of order milliseconds, whereas the redshift~$\znu$ varies on much longer timescales, typically months to years or longer. Consequently, the finite-period corrections are extremely small, and the observed timing modulation~$\ztau$ is accurately described by its leading term, namely the observed redshift~$\znu$ along a single geodesic. This leading description is the one conventionally employed in PTA observations \cite{NANO23ind12,NANO23new,NANO23gw,EPTA23UDM,EPTA23a,EPTA23C,IPTA23indBH}.

\section{Discussion}

Pulsar timing array (PTA) observations measure the arrival times of radio pulses from millisecond pulsars, while the theoretical description of the PTA observable~$\ztau$ in Eq.~\eqref{z2} is conventionally given by the Sachs--Wolfe formula~\cite{SAWO67} for the observed redshift~$\znu$ in Eq.~\eqref{z1}, which describes the change in photon frequency along a single geodesic. We have shown directly from the PTA timing observable that the two redshifts are exactly equivalent in the limit of vanishing proper-time separation between the emission events, without relying on a perturbative expansion of the spacetime geometry.

The reason this equivalence arises naturally in pulsar timing arrays is that the two radio pulses correspond to successive realizations of the same source--observer propagation link. As the pulse separation tends to zero, the second trajectory becomes the infinitesimal variation of the first, rather than an independent finite trajectory. The two-pulse timing observable can therefore be related exactly to the observed redshift evaluated along a single reference geodesic.

For a finite emission interval, the departure from this limit is controlled by the expansion
parameter~$\Delta\tau_\emi/T_z$ in Eq.~\eqref{finite-expansion}. Importantly, the smallness of this parameter is completely independent of the perturbative expansion of the spacetime geometry. Equation~\eqref{finite-average} is exact and follows solely from the smooth emission--observation map~$\mathcal{T}(\tau_\emi)$ in an arbitrary metric spacetime, while Eq.~\eqref{finite-expansion} is an expansion only in the emission interval~$\Delta\tau_\emi$. Neither expression requires an expansion of the spacetime geometry. Our result therefore
shows that the use of the single-geodesic observed redshift as the leading description of the PTA signal~$\ztau$ is \textit{not} restricted to first-order perturbation theory. At linear order, this redshift is given by the usual Sachs--Wolfe formula, while beyond linear order it can be evaluated to the corresponding perturbative order along the same reference geodesic. Equation~\eqref{finite-average} provides the corresponding finite-period relation at any perturbative order.

\bigskip
\acknowledgments
We acknowledge useful discussions with Philippe Jetzer. The work of MM was supported by IBS under the project code IBS-R018-D3.

\bibliographystyle{JHEP}
\bibliography{ms.bbl}

\end{document}